\documentclass[preprint,amsmath,amssymb,aps,floatfix,showpacs]{revtex4-1}
\usepackage{graphicx}
\usepackage{dcolumn}
\usepackage{bm}
\usepackage{braket} %Akbar Added
\usepackage{color} %Akbar Added
\usepackage{gensymb}

\begin{document}

%\preprint{APS/123-QED}

\title{Anisotropic Interactions in Electric Field Polarized Ultracold Rydberg Gases}

\author{Akbar Jahangiri}
\author{James P. Shaffer}
\affiliation{Homer L. Dodge Department of Physics and Astronomy,The University of Oklahoma, \\440 W. Brooks Street, Norman, OK 73019, USA}

%\affiliation{Quantum Valley Ideas Laboratories, 485 Wes Graham Way, Waterloo, ON, N2L 0A7 Canada}

\author{Lu\'{\i}s Felipe Gon\c{c}alves}
\author{Luis Gustavo Marcassa}
\affiliation{Instituto de Fı́\'{a}sica de S\~ao Carlos, Universidade de S\~ao Paulo, Caixa Postal 369, 13560-970, S\~ao Carlos, S\~ao Paulo, Brazil.}

\date{\today}

\begin{abstract}
We calculate pair potential curves for interacting Rydberg atoms in a constant electric field and use them to determine the effective $C_3$ dipole-dipole and $C_6$ van der Waals coefficients.  We compare the $C_3$ and $C_6$ with experiments where the angle of a polarizing electric field is varied with respect to the axis of a quasi-1-dimensional trap at ultracold temperatures. The dipoles produced via polarization of the atoms have an angular dependent dipole-dipole interaction. We focus on the interaction potential of two rubidium Rydberg atoms in $50S_{1/2}$ states in the blockade regime. For internuclear distances close to the blockade radius, $R_{bl} \approx 4 - 6\,\mu$m, molecular calculations are in much better agreement with experimental results than those based on the properties of single atoms and independent calculations of $C_3$ and $C_6$ which were used to analyze the original experiment. We find that the calculated $C_6$ coefficient is within $8\%$ of the experimental value while the $C_3$ coefficient is within $20\%$ of the experimental value.
\end{abstract}

%\pacs{36.90+f, 39.25+k, 32.10-f, 33.80.Rv}

\maketitle

\section{\label{sec:Introduction}Introduction}
Since the excited electron in a Rydberg atom is very loosely bound to its ionic core, Rydberg atoms exhibit very high polarizability. As a result, the interactions between Rydberg atoms can be particularly strong \cite{AMOP2014}. Rydberg atom interactions can be resources or liabilities depending on how the Rydberg atoms are employed. There have been many studies done on Rydberg atom interactions because of the wide range of applications where they are integral \cite{SpecIssue2015}. Amongst the many examples, ultracold Rydberg atoms are used in the study of quantum information \cite{Saffman2010,Saffman2016}, surfaces \cite{sedlace2016}, quantum optics \cite{Charles2016} and dipolar quantum gases \cite{Lahaye2009}. Phenomena such as the formation of novel types of molecules \cite{NatureComm}, collisions between ultracold Rydberg atoms, including the effect of electric fields \cite{Marcassa2009,Cabral2011}, and dipole-dipole and van der Waals interactions \cite{AMOP2014,Lahaye2013} have been investigated. The Rydberg atom blockade effect \cite{Lukin2001}, whose basis lies in the strength of the Rydberg atom interactions, has attracted widespread attention. Interactions between Rydberg atoms also play an important role in investigating the intrinsic properties of cold matter \cite{SpecIssue2015} and in electric field sensing \cite{Fan2015}.

%==============
\begin{figure}
	\includegraphics[scale=0.42]{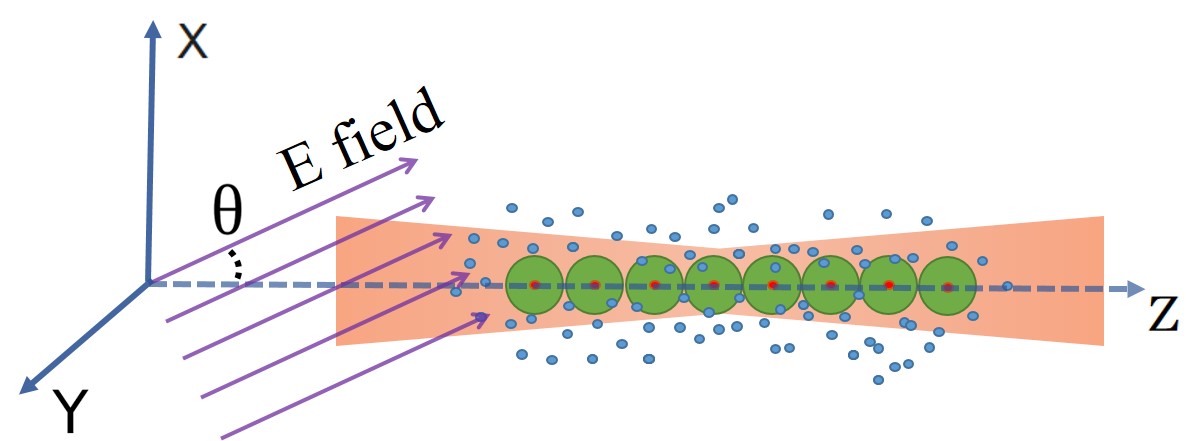}% Here is how to import EPS art
	\caption{(Color online) One dimensional trapped atomic sample in an electric field. The figure illustrates the one dimensional distribution of 'super atoms' excited due to Rydberg atom blockade along the axis of the dipole trap. The figure defines the tilt angle of the electric field relative to the trap axis, $\theta$.}\label{fig:1dsample}
\end{figure}
%==============

When a Rydberg atom is excited, its presence will change the energy required for a ground state atom in its vicinity to be excited to the Rydberg state. This change in energy creates a density limitation for the Rydberg atomic population due to the strong interaction between two nearby Rydberg atoms and is referred to as the Rydberg blockade effect \cite{Lukin2001}. The illustration in Fig.~\ref{fig:1dsample} depicts blockade regions around Rydberg atoms in an atomic gas. The Rydberg atom blockade phenomena enables the control of atomic excitation and de-excitation in a gas. With strongly interacting Rydberg atoms, single atom excitation states are no longer degenerate. One way of changing the Rydberg atom interaction strength to control the blockade effect is to change the orientation of an applied electric field~\cite{Marcassa2016}.

If Rydberg atoms are positioned far from each other; $R$ on the order of a few microns, depending on the states involved, the leading order term of the interaction is a dipole-dipole interaction which leads to van der Waals, $\propto R^{-6}$, and resonant dipole, $\propto R^{-3}$, potential energy curves. Higher order interactions and electric fields can also be important. For example, quadrupole interactions can cause state mixing as well as small but experimentally significant energy shifts \cite{Arne2006,Booth2010,Pillet2014}. Electric fields cause the atomic state energies of the Rydberg atoms to shift, as well as mix zero-field atomic states. The electric field polarizes the atoms inducing atomic dipoles. In some applications, the background electric field can be used to advantage. One advantage of applying a controlled background electric field is the possibility of tuning interaction potentials. As an example, one can create potential wells that can hold many bound states of Rydberg atom pairs \cite{Arne2007,Overstreet2009}. In many Rydberg atom experiments, all these effects are important to consider in order to explain experimental results quantitatively, as none of the effects are generally independent of one another.

%\begin{figure}
%\includegraphics[scale=0.25]{blockade.png}
%\caption{\label{fig:blockade}(Color) Blockade radius created around Rydberg atoms. Blockade regions are created because of interaction between Rydberg atoms. In blockade regions ground state atoms can not be excited to the Rydberg state.}
%\end{figure}

In this paper, we use calculations of Rydberg atom potential energy curves and surfaces to explain results obtained in Ref.~\cite{Marcassa2016} quantitatively. In that paper, it was shown that one can enhance or suppress the blockade effect in a quasi 1-dimensional sample of rubidium atoms by tilting an electric field by an angle, $\theta$, with respect to the axis of the sample, Fig.~\ref{fig:1dsample}. Tilting the electric field with respect to the trap axis can change the orientation of the electric field induced dipoles in the 1-dimensional sample of Rydberg atoms, hence changing the strength of the interactions along the axis of the trap \cite{Noel2004}. We performed a series of pair potential calculations for the case of the $50S_{1/2}+50S_{1/2}$ rubidium Rydberg atom pair state. Single-atom Stark calculations \cite{Matthias2013,Pohl2011} were used in the original work to estimate the dipole-dipole interaction coefficient $C_3$ arising from the electric field induced dipoles. The $C_6$ coefficient was obtained using a perturbative calculation. The results of this theoretical approach lead to a $C_3$ that deviated from the experimental results by a factor exceeding 7 \cite{Marcassa2016}. Our calculations, which take into account the leading multi-pole interactions between Rydberg atoms and applied electric field, show better agreement with the experiment. Our $C_3$ coefficient differs by $20\%$ from the one obtained by fitting the experimental data,  which is within experimental error for this type of measurement. In addition, our $C_6$  coefficient is in better agreement with the experimental values, differing only by $\sim 8\%$ rather than $\sim 17\%$.

Our results demonstrate that the $C_6$ and $C_3$ coefficients are not independent of one another and the asymptotic approximation, traditionally used to calculate long range potentials for ground state atoms, is frequently invalid for Rydberg atoms over the most important internuclear separations. A straightforward way to see the interdependence is to notice that as the 2 atoms approach each other the van der Waals interaction hybridizes the orbitals of the atoms which changes the polarizability of each atom as R changes, resulting in what is effectively a R-dependent electric field induced dipole-dipole potential. The dipole-dipole potential that would result from 2 otherwise non-interacting Rydberg atoms is not the same as when the atoms are interacting via another strong multipolar interaction. Practically, this means that one must exercise care when using asymptotic polarizabilities to directly determine the $C_3$ coefficient resulting from electric field induced atomic dipoles. In such cases, matrix diagonalization, rather than perturbation theory, is generally required to calculate Rydberg atom interactions accurately.

\section{\label{sec:Experimental_details}Experimental}

In the experiment, a tightly confined atomic sample is prepared and held in a quasi-electrostatic trap (QUEST). The QUEST produced a non-polarized, approximately one-dimensional atomic sample, Fig.~\ref{fig:1dsample}, with $\sim10^6$ atoms at a peak density of $\approx 10^{12}\,$cm$^{-3}$ at a temperature of $60\, \mu$K. This information along with the excitation laser intensities is used to find the average internuclear distance between Rydberg atoms, $\sim 4 - 6\,\mu$m, and ground state atoms.

The QUEST is created by a linear polarized $10.6\,\mu$m CO$_2$ laser. The power of the CO$_2$ laser was $80$W. To load the trap, the CO$_2$ laser was focused into a magneto-optical trap. The CO$_2$ beam waist is $15\,\mu$m \cite{Marcassa2016}. The important experimental parameter is the angle between the QUEST axis and a constant, external electric field that is applied to the atomic sample, $\theta$ in Fig.~\ref{fig:1dsample}. The details of the experimental setup can be found in Ref.~\cite{Kondo2014,Kondo2016,Marcassa2016}.
%==============

The first step of the experiment was to obtain a Stark shift spectrum at the highest achievable density in order to find the electric field strength at which the dipole-dipole and van der Waals interactions are comparable (Fig. 1(b) in reference~\cite{Marcassa2016}), $2370\,$mV$\,$cm$^{-1}$. Rydberg atoms are excited using a 2-photon process consisting of the absorption of a $780\,$nm photon and a tunable photon at $\sim 480\,$nm. To experimentally determine the interaction potentials between the $50S_{1/2}$ Rydberg atoms, the population of the initially excited state was measured as a function of ground state atomic density for several $\theta$ at an electric field amplitude of $2370\,$mV$\,$cm$^{-1}$, at a single atom excitation laser detuning of $\Delta_{480} =-151\,$MHz (Fig. 2 in reference~\cite{Marcassa2016}). The experimental results clearly show that the blockade effect depends on $\theta$. The work also showed that the measurement performed at the magic angle is very similar to the measurement performed at zero field, suggesting an electric field induced dipole-dipole interaction as well as the 1-dimensional character of the atomic sample are important to interpret the results.
%==============
%\begin{figure}
%	\includegraphics[scale=0.3]{luis-stark.png}% Here is how to import EPS art
%	\caption{\label{fig:luis-stark}(Color) (a) Normalized $50S_{1/2}$ population as a function of the dc electric field for $\Delta_{480} =-151 $ MHz. (b) The Experimental Stark spectrum showing the avoided crossings of the $50S_{1/2}$ state with the
%		manifold hydrogenic lines~\cite{luis}.
%	}
%\end{figure}
%==============

Using the results of the measurements in Ref.~\cite{Marcassa2016}, $C_3$ and $C_6$ coefficients were extracted. Using the $50S_{1/2}$ rubidium Rydberg state population measured as a function of ground state atomic density for several $\theta$, a classical hard-sphere model in the steady state was applied \cite{Matthias2013,Pohl2011} to determine $C_3$ and C$_6$,
\begin{equation}\label{hardsphere}
\begin{split}
& \frac{\rho_{Ryd}(\theta)}{\rho} = \\ & \frac{3 - \frac{3}{2} \sqrt{\frac{4}{9} \pi^2 \rho^2 (\sqrt{4 C_6 + C_3^2 P(\theta)^2} + C_3 P(\theta))^2 +4}} {2 \pi \rho (\sqrt{ 4 C_6 + C_3^2 P(\theta)^2} + C_3 P(\theta))} +\frac{1}{2}
\end{split}
\end{equation}
where
\begin{equation}
P(\theta)= 1 - 3 \,\mathrm{cos}^2(\theta).
\end{equation}
$\rho$ is the ground state density and $\rho_{Ryd}(\theta)$ is the angle dependent Rydberg density.  The hard-sphere model treats Rydberg atoms as hard spheres so that Rydberg excitations are not allowed to overlap. The radius used to obtain the blockade volume, $V_{bl}$, is the blockade radius, $R_{bl}$, where,
\begin{equation}
V_{bl} = \frac{4 \pi}{3} R_{bl}^3.
\end{equation}
Under these conditions, the experiment determined $C_6^{exp} = 18 418\,$MHz$\,\mu$m$^6$ and $C_3^{exp} = 99.74\,$MHz$\,\mu$m$^3$. The van der Waals interaction parameter, $C_6$, is close to the theoretical one for the $50S_{1/2}$ state, $C_6^
{ST} = 15 296\,$MHz$\,\mu$m$^6$ obtained perturbatively \cite{Raithel2007}. However, the dipole-dipole interaction parameter is seven times larger than the theoretical one, $C_3^{ST}= 14.375\,$MHz$\mu$m$^3$, obtained from a single-atom Stark calculation. The discrepancy is due to that fact that these methods of calculation do not fully explain the complex nature of the interaction between two Rydberg atoms. Because high angular momentum states are energetically nearby, ($l_{max}=n-1$), the multilevel character of the interaction needs to be included to obtain accurate potential curves \cite{Marcassa2016,Kondo2016}. The multitude of quasi-resonant interactions makes the asymptotic calculation of the interaction invalid over a large and relevant range of internuclear separations, particularly near the blockade radius. The interaction of many levels changes the polarizability of the atoms as a function of $R$, making the electric field induced dipole-dipole interaction different from what it would be if the asymptotic polarizabilities were used to calculate its magnitude.

\section{\label{sec:Theory_details}Theory}

To achieve control over Rydberg atom blockade, the orientation of the induced atomic dipole moments was changed in Ref.~\cite{Marcassa2016} by changing the external dc electric field relative to the long axis of the trap, $\theta$. In the presence of an external electric field, the interaction potential changes to an angular dependent potential because the atoms become polarized~\cite{Marcassa2016,Noel2004}. Note that the potentials corresponding to individual magnetic sub-levels are also angularly dependent, but if there is no field to provide an orientation in space there is no angular dependence, as there exists an isotropic, degenerate superposition of magnetic sub-levels. The orientational field can be supplied by a continuous field, as primarily addressed in this work, a laser through its polarization, or by other interactions in the molecular frame. In this work, we investigate asymptotic S-states which are described approximately by isotropic van der Waals potentials when the atoms are far apart. Intuitively, under the experimental conditions, the interaction potential consists of two pieces, the van der Waals and the dipole-dipole potentials. The effective interaction potential, adopting this approximation, can be written as
\begin{equation}
\label{eq:veff}
V_{eff}(R,\theta)=\frac{C_6}{R^6} + \frac{C_3}{R^3} [1-3 \cos^2(\theta)].
\end{equation}
In this equation, $C_3$ is the dipole-dipole interaction parameter, resulting from the polarization of the atoms in the external electric field, and $C_6$ is the van der Waals coefficient. The dipole-dipole potential in the present case can be written as a semi-classical expression,
%==========
\begin{equation}
\label{eq:dipole-dipole}
V(R) = \frac{\mu^2}{4 \pi \epsilon_0 R^3} (\cos(\theta_{12})-3\cos(\theta_1)\cos(\theta_2)).
\end{equation}
%==========
$\mu$ is the electric field induced permanent dipole of the atoms which we assume to be in the same atomic state. $\theta_{12}$ is the angle between two dipoles, and $\theta_1$ ($\theta_2$) is the angle between the dipole of atom 1 (atom 2) and the internuclear axis. Since the permanent dipoles are produced as a result of an external electric field and our sample in the experiment is a quasi 1-dimensional sample, the dipole orientations are always in the direction of the external electric field. Thus, $\theta_{12}$ is always zero and $\theta_1=\theta_2$. The simplified version of Eq.~\ref{eq:dipole-dipole} can be written as
%=============
\begin{equation}
\label{eq:dipole-dipole-2}
V(R) = \frac{\mu^2}{4 \pi \epsilon_0 R^3} (1-3\cos^2(\theta)).
\end{equation}
%=============
The prefactor in this expression is the $C_3$ coefficient. The angular dependence in the potential energy is given by $\theta$.

These expressions are an intuitive way to think about what is happening in the experiment. In fact, we later show by calculating the potential energies according to the method found in Ref.~\cite{Arne2006} that the dominant interactions are a field induced dipole-dipole interaction and a van der Waals interaction. Consistent with the experiment, the magnitudes of these interactions, particularly the electric field induced dipole-dipole interaction, cannot be accurately calculated without considering both interactions simultaneously, i.e. these interactions are not independent of each other and should be diagonalized together. The issue with using these expressions to calculate the angular dependence of the interactions can be understood if we write down an expression for $C_3$,
%=================
\begin{eqnarray}
\label{eq:one14}
C_3 =\frac{e^2 a_0^2}{4 \pi \epsilon_0} && \braket{\widetilde{50S_{1/2}50S_{1/2}}|r_1|\widetilde{50S_{1/2}50S_{1/2}}} \nonumber\\ \times&& \braket{\widetilde{50S_{1/2}50S_{1/2}}|r_2|\widetilde{50S_{1/2}50S_{1/2}}}.
\end{eqnarray}
%=================
Here, $\braket{\widetilde{50S_{1/2}50S_{1/2}}|r_1|\widetilde{50S_{1/2}50S_{1/2}}}$ is the dipole moment of atom 1 in an external electric field while the same expression with $r_2$ gives the dipole moment of the second atom. In the dipole moment expressions, $r_1$ only acts on electron 1 and $r_2$ acts on electron 2. $e$ is the electron charge and $a_0$ is the Bohr radius. Here we take the most plausible approach of using the molecular wavefunction in the background electric field, indicated by the tilde and the relevant Rydberg pair state $50S_{1/2}+50S_{1/2}$ \cite{Arne2006}. However, the problem with the expression is that it is clearly dependent on $R$ due to the van der Waals interaction. Although this is a straightforward argument, it effectively illustrates why calculating the dipole moments independently using the asymptotic atomic states can result in a poor approximation to the angular dependent interaction potentials.

%==============
\begin{figure}
    \begin{center}
	\includegraphics[scale=0.34]{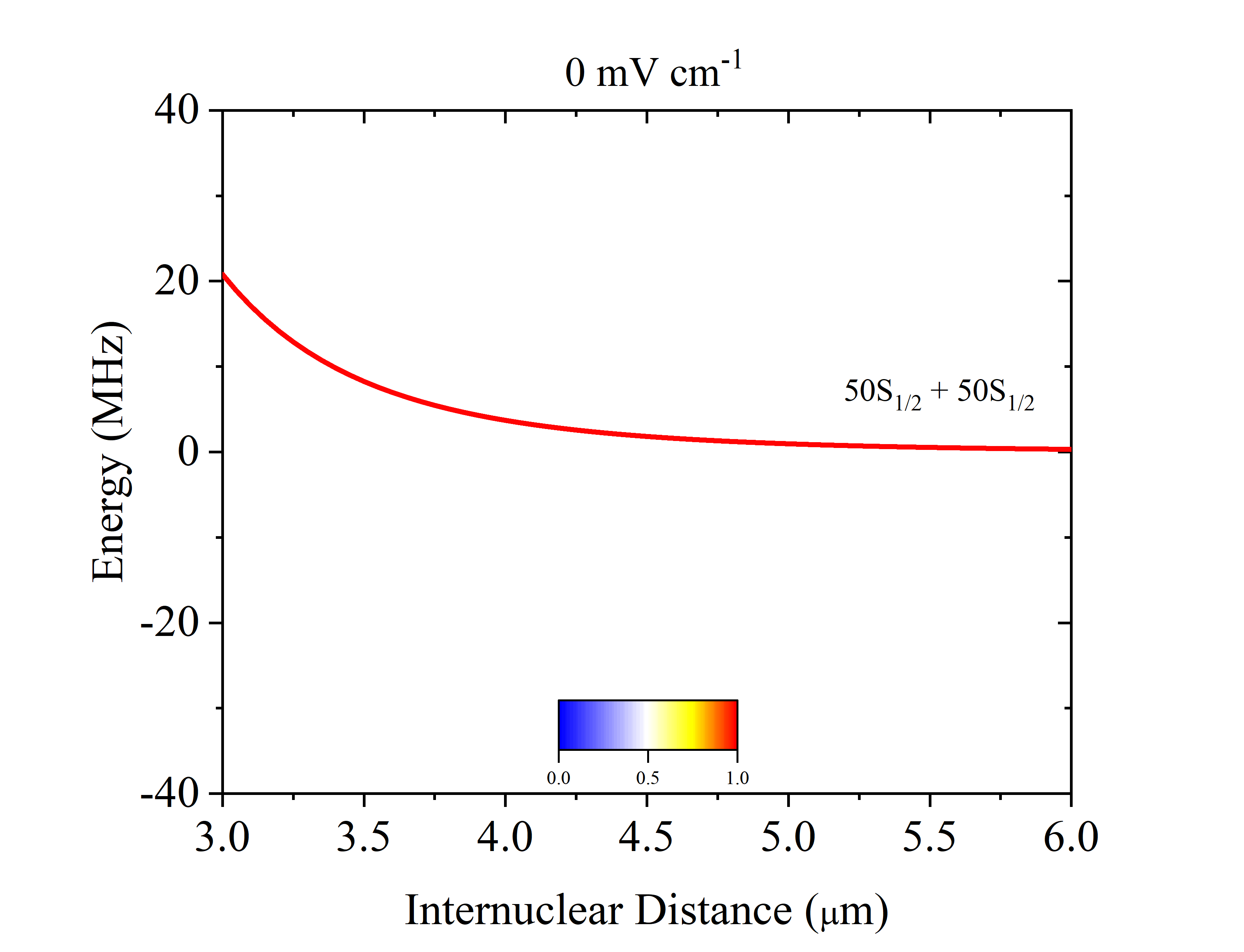}% Here is how to import EPS art
    \end{center}
	\caption{\label{fig:zero-e-field}(Color online) Interaction potential curve of the $50S_{1/2}+50S_{1/2}$ in zero electric field. Here the source of interactions is mainly of the van der Waals type. The population of $50S_{1/2}+50S_{1/2}$ is large for distances larger than $1.5\,\mu$m. The magnitude of the $50S_{1/2}$ component is shown for comparison with the potentials at $2370\,$mV$\,$cm$^{-1}$. The color indicates the fraction of $50S_{1/2}+50S_{1/2}$ in the state. As can be observed in the plot, the state is almost entirely $50S_{1/2}+50S_{1/2}$.
	}
\end{figure}
%==============

%==============
%\begin{figure}
%	\includegraphics[scale=0.3]{dipole_dipole.png}% Here is how to import EPS art
%	\caption{\label{fig:dipole_dipole}(Color) Direction of electric field and induced dipole moments in two Rydberg atoms. The black arrow passing through both atoms is the %internuclear axis. (a) The electric field is parallel to the direction of internuclear axis of the Rydberg atoms, $\theta=0^o$. (b) The direction of electric field is %perpendicular to the  internuclear axis of the Rydberg atoms, $\theta=90^o$.
%	}
%\end{figure}
%==============

Based on the experimental work, we performed a series of pair potential calculations for rubidium in the $50S_{1/2}+50S_{1/2}$ quantum state for various electric field $\theta$. By simultaneously considering multipolar interaction terms up to the quadrupole between Rydberg atoms, we have found a more accurate result for the dipole-dipole coefficient $C_3$, and van der Waals coefficient, $C_6$, by fitting our results to the model potential. These fits demonstrate that qualitatively the interaction studied in the experiment is dominated by the isotropic van der Waals and anisotropic electric field induced dipole-dipole potentials. Quantitatively, the fitted $C_6$ and $C_3$ agreed with the experimental results much more closely than the original estimates based on the asymptotic atomic wavefunctions.

\begin{figure}
    \begin{center}
	\includegraphics[scale=0.36]{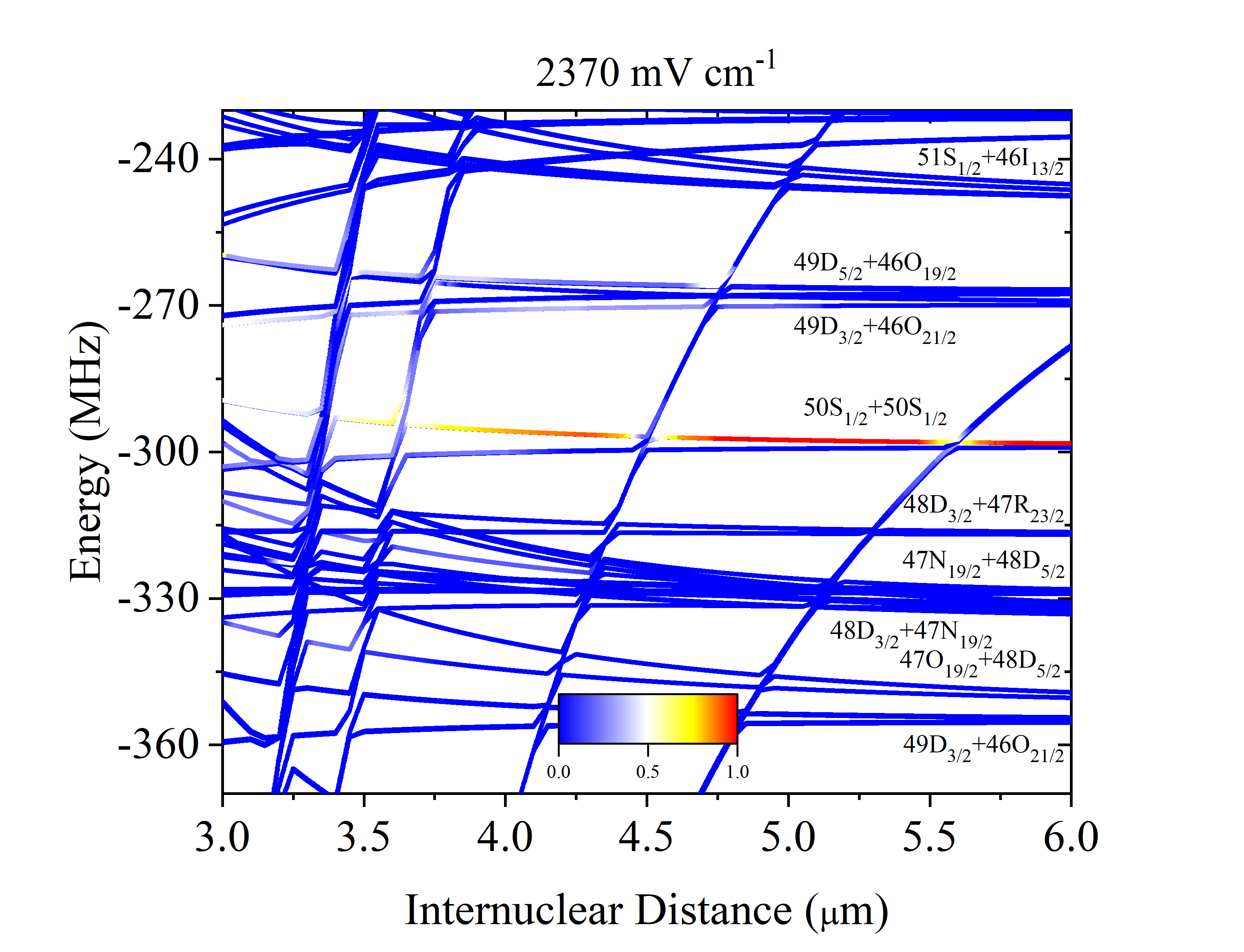}
    \end{center}
	%    \captionsetup{format=hang}
	\caption{(Color online) Pair potential for $50S_{1/2}+50S_{1/2}$ in an electric field of $2370\,$mV$\,$cm$^{-1}$ for $\theta = 0\,$degrees. The darker red color shows higher $50S_{1/2}+50S_{1/2}$ population for different internuclear distances for a constant electric field. The potential energy curves shown in the figure are very different from those with 0 applied electric field because the Stark effect causes some of the n=47 Stark manifold states to intersect the $50S_{1/2} +50S_{1/2}$ curves. This is essentially why the electric field induced dipole-dipole interaction is stronger than first calculated. The interplay between the electric field and the Rydberg atom interactions is complicated because it is highly multilevel in nature for this particular case. There is a small, $\sim 1\,$MHz per atom, energy difference between the theoretical calculation and the experiment which is within the error of the frequency reference used for the experiment. The zero of the energy scale is set at the field free $50S_{1/2}+50S_{1/2}$ asymptote.}
	\label{fig:potential-pairs-color}
\end{figure}

When the electric field is held at zero, the Rydberg atoms do not have permanent dipoles and the interaction between Rydberg atoms at the blockade radius is predominantly a van der Waals interaction. Fig.~\ref{fig:zero-e-field} shows the potential curves along with the population of the $50S_{1/2}+50S_{1/2}$ quantum state for internuclear distances $R=3\,\mu$m to $R=6\,\mu$m. The potential in Fig.~\ref{fig:zero-e-field} is weakly repulsive in zero electric field. For the electric field used in the experiment there are substantial changes to the interaction potentials as shown in Fig.~\ref{fig:potential-pairs-color}. Most notably, the strength of the $50S_{1/2}+50S_{1/2}$ potential changes in an electric field of $2370\,$mV$\,$cm$^{-1}$.

%=============
%==============

We first kept $\theta$ fixed at zero degrees and examined the interaction potential for different internuclear distances. The result of the pair potential calculations for an electric field of $2370\,$mV$\,$cm$^{-1}$ for $\theta = 0$ is shown in Fig.~\ref{fig:potential-pairs-color}. We also calculated the amplitude of $50S_{1/2}+50S_{1/2}$ as a function of internuclear distance for a background electric field of $2370\,$mV$\,$cm$^{-1}$. The amplitude allowed us to verify the potential that was excited in the experiment. We calculated the population by first obtaining the asymptotic wavefunctions for the state of interest, $50S_{1/2}+50S_{1/2}$, and nearby states. When the atoms are interacting, the wavefunction of the system can be written as a superposition of asymptotic atom pair wavefunctions,
\begin{eqnarray}
\label{eq:one13}
\ket{\widetilde{50S_{1/2}50S_{1/2}}}=\alpha(R) {\ket{50S_{1/2} 50S_{1/2}}} \nonumber\\ +\sum_{n,l,n^\prime,l^\prime} \beta_{n,l,n^{\prime},l^\prime}(R) \ket{n l \, n^{\prime} l^{\prime}}.
\end{eqnarray}
$\ket{\widetilde{50S_{1/2}50S_{1/2}}}$ consists of mixture of all states that have interactions with $50S_{1/2}+50S_{1/2}$ or are coupled by the external electric field. We calculate the probability amplitude of $50S_{1/2}+50S_{1/2}$ using $\braket{50S_{1/2}50S_{1/2}|\widetilde{50S_{1/2}50S_{1/2}}}$ for each internuclear distance, $R$. $\ket{\widetilde{50S_{1/2}50S_{1/2}}}$ is the state that carries mostly $50S_{1/2}+50S_{1/2}$ character, which is shown in red in Fig.~\ref{fig:potential-pairs-color}. The higher shading of the red color shows a larger component of $50S_{1/2}+50S_{1/2}$. The mixing between $50S_{1/2}+50S_{1/2}$ and neighboring states can be seen as the internuclear distance between the Rydberg atoms changes. Fig.~\ref{fig:potential-pairs-color} is more complicated than Fig.~\ref{fig:zero-e-field}, because the electric field that polarizes the atoms is large enough so that the nearby n=47 Stark manifold is driven into the $50S_{1/2}+50S_{1/2}$ state. The fact that the state of interest interacts with high angular momentum states results in a larger polarizability and therefore a larger electric field induced dipole-dipole interaction. In Fig.~\ref{fig:potential-pairs-color} the interaction is dominated by the dipole-dipole potential since it is much stronger at $\theta = 0$ than the van der Waals contribution.

%==============
\begin{figure}[t!]
	\includegraphics[scale=0.35]{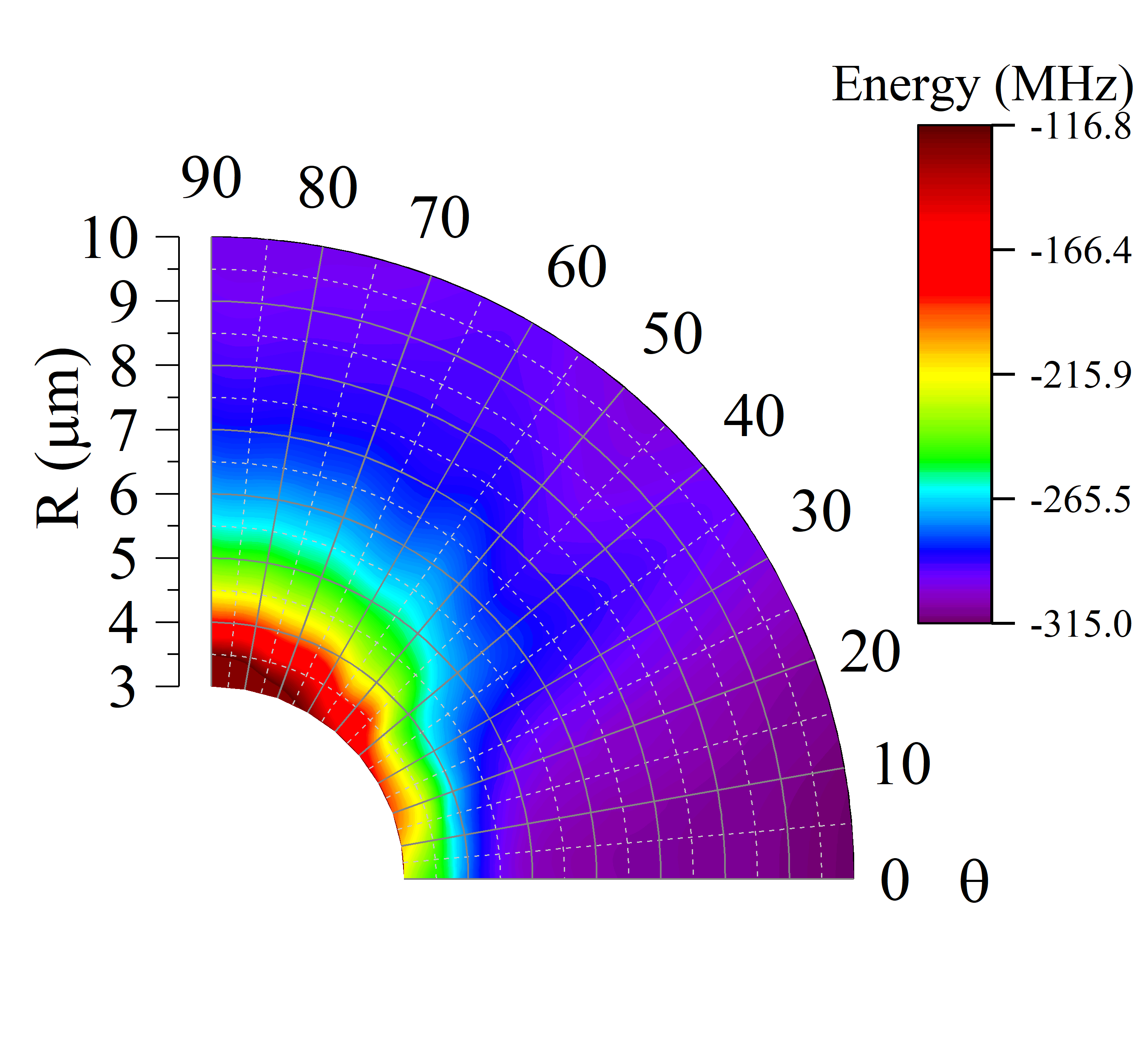}
	\caption{\label{fig:angular_energy}(Color online) Tilting the direction of the electric field changes the potential energy surface of the interacting Rydberg atoms each in the $50S_{1/2}$ state at an electric field of $2380\,$mV$\,$cm$^{-1}$. By changing the direction of the electric field, the magnitude of the cos($\theta$) term in $V_{eff}$ changes. Both the van der Waals and dipole-dipole parts of the potential energy surface are shown. Zero energy is chosen to correspond to the zero electric field $50S_{1/2} +50S_{1/2}$ asymptote.
	}
\end{figure}

Performing anisotropic calculations for various background electric field $\theta$ enables us to study the angular dependence of the dipole-dipole interaction while taking account of the van der Waals interaction. The interaction between the two Rydberg atoms is seen to depend on the direction of the applied background electric field in Fig.~\ref{fig:angular_energy}.
The plot shows the full interaction potential, including both the dipole-dipole and van der Waals contribution, of $50S_{1/2}+50S_{1/2}$ rubidium Rydberg atoms in an electric field of $2380\,$mV$\,$cm$^{-1}$. Calculating the potential at the magic angle which satisfies the $P_2(\mathrm{cos}\,\theta)=0$, where the $P_2$ is the Legendre polynomial of second order, results in an interaction potential that just contains the van der Waals potential, $\theta = 54.7^{\degree}$. Subtracting the van der Waals potential from the full potential, leaves the dipole-dipole interaction potential since the model van der Waals potential is isotropic, approximated as a pure S-state, and the dipole-dipole interaction described here is diagonal, i.e. the dipole-dipole interaction does not couple atoms in different states since these are 'static' dipoles formed by the polarization induced by the external electric field in our approximation. The potentials were fit in the region between $3\,\mu$m to $6\,\mu$m because this region spanned the blockade radius, Fig.~\ref{fig:potential-pairs-color}.

%==============
\begin{figure}[t!]
	\includegraphics[scale=0.33]{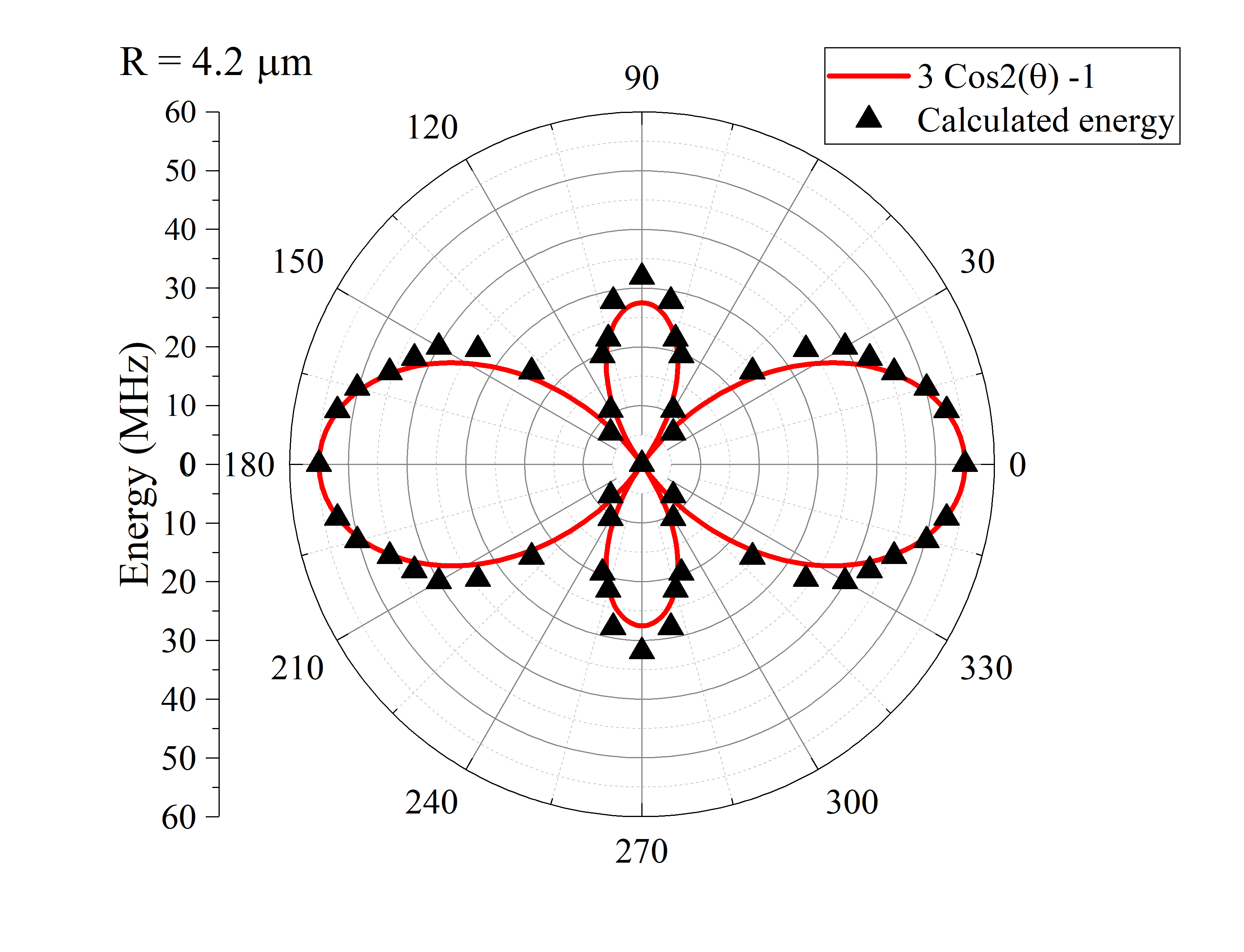}
	\caption{\label{fig:dipolar}(Color online) The magnitude of the dipole-dipole interaction part of the potential energy at $4.2\,\mu$m for an electric field of 2380$\,$mV$\,$cm$^{-1}$. The diagram shows that the angular dependence of the potential energy surface is well-represented by a dipole-dipole interaction. The potential is attractive along $\theta = 0$ degrees and repulsive along $\theta = 90$ degrees.
	}
\end{figure}
%==============

Fig.~\ref{fig:dipolar} shows the magnitude of the static dipole-dipole interaction potentials for $R = 4.2\,\mu$m for various angles. This figure demonstrates that the remaining interaction potential is well characterized as a dipole-dipole interaction around the blockade radius, as argued. Changing the orientation or magnitude of the constant external electric field changes the interaction strength between the Rydberg atoms. For example, in Fig.~\ref{fig:angular_energy}, when $\theta$ is zero, we have maximum attraction between the two induced dipoles, because the opposite charges of the dipoles are closer to each other. Changing the orientation of the external electric field to $\theta=90^o$ will result in maximum repulsion between the two dipoles, since the orientation of induced dipoles are perpendicular to the internuclear axis which causes like charged poles of the dipoles to be closer to each other. The effect is also observed in Fig.~\ref{fig:angular_energy} where the potential is slightly attractive along $\theta = 0$ degrees but is repulsive at short range and more strongly repulsive along $\theta = 90$ degrees.

If we use the full potential calculations to obtain the $C_3$ coefficient around the blockade radius, $R = 3- 6\,\mu$m, by fitting the model potential to the calculations shown in Fig.~\ref{fig:potential-pairs-color} and Fig.~\ref{fig:angular_energy}, we obtain $C_3\sim 120\,$MHz$\,\mu$m$^3$, which is in reasonable agreement with the experiment, given the prior estimates found in Ref.~\cite{Marcassa2016}. The theoretical value for $C_3$ is approximately 20\% different from the experimentally obtained result of $C^{exp}_3=99.7\,$MHz$\,\mu$m$^3$. The value obtained by single-atom Stark calculations for $C_3$ is $14.375\,$MHz$\,\mu$m$^3$, which is approximately 7 times smaller than the experimental value. The $C_6$ coefficient is also obtained by fitting, Fig.~\ref{fig:potential-pairs-color}, the calculated potential to the model potential. Using our method the van der Waals coefficient is found to be $C_6 \sim 19970\,$MHz$\,\mu$m$^6$ which is in better agreement with the experimental value than the estimates in Ref.~\cite{Marcassa2016}. The fact that $C_3$ is larger than estimated is due to the fact that the electric field that is applied in the experiment is large enough to shift the $50S_{1/2}+50S_{1/2}$ state into the nearby n=47 Stark manifold. The mixing with the high angular momentum states is important and demonstrates the multilevel nature of the interatomic forces near the blockade radius in this case, consistent with the experiment.

\begin{figure}[t!]
	\includegraphics[scale=0.33]{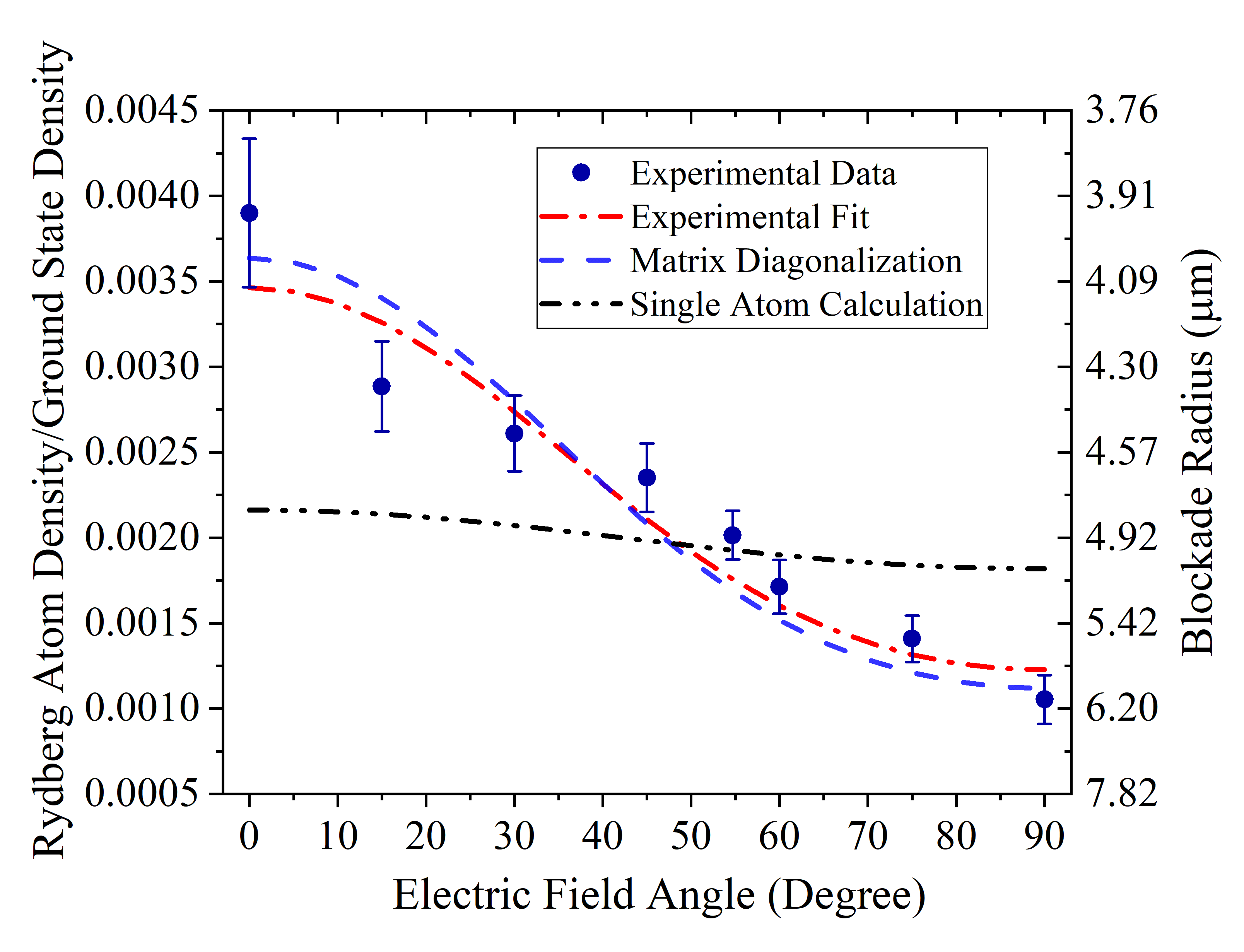}
	\caption{\label{fig:magic_angle}(Color online) This figure plots the number of Rydberg atoms per ground state atom that are excited at different electric field angles, $\theta$. The points are the data from Ref.~\cite{Marcassa2016}. The red dot-dashed line shows the fit from which $C_3$ and $C_6$ were obtained from the experimental data using a nonlinear least squares fit. The blue dashed line shows the results of the calculation presented in the paper. There are no fitting parameters used to plot the blue dashed line, only the experimental measured density is used for the plot.  The black dot-dot-dashed line is the estimated theoretical result presented originally in Ref.~\cite{Marcassa2016}. The agreement between the experimental results and the full atom-pair calculation is similar to the nonlinear least squares fit of the experimental data. The right hand y-axis labels are how the blockade radius changes as a function of $\theta$ based on the assumption that the peak ground state density is $10^{12}\,$cm$^{-3}$. It is a straightforward nonlinear scaling based on how many ground state atoms fit into a blockade sphere at uniform density.}
\end{figure}

Figure~\ref{fig:magic_angle} shows the number of $50S_{1/2}$ atoms per ground state atom as well as the blockade radius as a function of $\theta$. The fit to the hard sphere model that was used to extract $C_6^{exp}$ and $C_3^{exp}$ is also shown in Fig.~\ref{fig:magic_angle} as is the graph of the hard sphere model based on our calculation of the electric field dependent interaction potential. Fig.~\ref{fig:magic_angle} is the central result of the paper, showing the relationship between the experimental data and calculations. The lines in the figure are obtained by considering the Rydberg density obtained using the hard sphere model, Eqn.~\ref{hardsphere}. Although the hard sphere model assumes a repulsive potential, we use it for our analysis even at $\theta = 0\,$degrees where the potential becomes slightly attractive. The attractive part of the potential is $< 1\,$MHz, which is the linewidth of the lasers. The blockade radius changes from $4\,\mu$m to $6\,\mu$m as $\theta$ sweeps from 0 to 90 degrees. There are around $\sim 250-1000$ ground state atoms in a blockade volume. At $\theta = 0\,$degrees, where the dipoles are aligned head to tail, the blockade radius is at a minimum since the dipole-dipole potential is attractive while the van der Waals potential is repulsive. When $\theta =90\,$degrees, the potentials have the same sign and the blockade radius increases.

Fig.~\ref{fig:magic_angle} shows that the angular dependence of the actual potential is well described by the calculational approach of Ref.~\cite{Arne2006}. The application of the electric field to polarize the atoms clearly has an effect on the blockade radius - it is changed by $\pm 20\%$ when $\theta$ changes from 0 to 90 degrees. The theory curve in Fig.~\ref{fig:magic_angle} actually reproduces the data as well as the least squares fit over the extent of the range of $\theta$ if the clear systematic error in the experiment is taken into account. The clear outlying point at 15$\,$degrees probably has perturbed the least squares fit enough to shift it somewhat. The theory curve that we have calculated here, shown in Fig.~\ref{fig:magic_angle}, has no adjustable parameters. Only our fit of the calculation and the experimentally measured density are needed to calculate the curve. Although the fit of $V_{eff}$ and our calculation fit the data well, discrepancies between the data and experiment could arise because we are using a model to calculate the curves in Fig.~\ref{fig:magic_angle}. There are still higher order interactions and state mixing, so the splitting of the potential into one with dipolar angular dependence and an isotropic piece is not exact. The benefit of the simple picture presented in this paper outweighs the value of rigorously or semi-rigorously modifying $V_{eff}$. Including higher orders of interaction complicates the form of the interaction potentials. The experimental data and the degree to which it agrees with our approximation justifies our approach.

%===========================================
\section{\label{sec:Conclusion}Conclusion}

%==============

We theoretically calculated pair potentials for $50S_{1/2}+50S_{1/2}$ rubidium atoms in a constant electric field and used them to interpret experiments in a 1-dimensional sample of ultracold rubidium atoms in the blockade regime \cite{Marcassa2016}. We determined $C_3$ and $C_6$ coefficients in the effective potential relation, $V_{eff}$, by fitting our calculations in the vicinity of the blockade radius to $V_{eff}$ and compared them to the experimental results. We showed that our results are in better agreement with the experiment than estimates that treat these interactions independently and use asymptotic properties to calculate $C_3$ and perturbation theory to calculate $C_6$. The $C_3$ coefficient that we calculated is only $\sim 20\%$ larger than the value measured in the experiment at $R_{bl}$. The single-atom Stark calculation estimate originally used to interpret the experiment is 7 times smaller than the experimental value at $R_{bl}$. The $C_6$ coefficient that we calculated is also in better agreement with the experimental value when compared to the one originally used to interpret the results. Our $C_6$ is within $\sim 8\%$ of the measured value while the prior calculations are within $\sim 17\%$. Our anisotropic calculations reveal the importance of the static dipole-dipole interaction in the Rydberg blockaded sample, in further agreement with the conclusions of Ref.~\cite{Marcassa2016}. Although, one could completely discard $V_{eff}$, it's form is useful for interpreting how the atoms in the trap are interacting and Fig.~\ref{fig:angular_energy} and Fig.~\ref{fig:dipolar} show that the effective potential captures the essential physics. The interaction between two polarized Rydberg atoms is more complicated than what is considered in $V_{eff}$, especially at $R\sim R_{bl}$, but the $R$ and $\theta$ dependence is essentially correct. Experiments involving quantum-level dynamics and dense samples of blockaded atoms are areas where these results may be most relevant, particularly if electric fields are used to control the interactions between the atoms.

\section{Acknowledgements}
This work is supported by grants 2011/22309-8 and 2013/02816-8 from the S\~{a}o Paulo Research Foundation (FAPESP), Air Force Office of Scientific Research grant FA9550-16-1-0343 and CNPq. J.P.S and A.J. acknowledge support from the Air Force Office of Scientific Research grant FA9550-12-1-0282 and the National Science Foundation research grant PHY-1607296.

\bibliography{Anisotropic5}

%\begin{thebibliography}{63}

\end{document}